\documentclass{elsart}

 \usepackage{epsfig}

\usepackage{amssymb}

\newcommand{\bx}{{\bf x}}
\newcommand{\by}{{\bf y}}
\newcommand{\bu}{{\bf u}}
\newcommand{\bv}{{\bf v}}
\newcommand{\bk}{{\bf k}}
\newcommand{\bq}{{\bf q}}
\newcommand{\HH}{{\mathcal H}}
\newcommand{\BE}{\begin{equation}}
\newcommand{\EE}{\end{equation}}
\newcommand{\BA}{\begin{eqnarray}}
\newcommand{\EA}{\end{eqnarray}}

\begin{document}

\begin{frontmatter}


\title{Fluctuations impact on a pattern-forming model of population dynamics with non-local
interactions}

\author{Crist\'obal L\'{o}pez\thanksref{email1}}
\thanks[email1]{\tt clopez@imedea.uib.es}
\author{and Emilio Hern\'andez-Garc\'\i a\thanksref{email2}}
\thanks[email2]{\tt emilio@imedea.uib.es}

\address{Instituto
Mediterr\'aneo de Estudios Avanzados IMEDEA (CSIC-UIB)
\thanksref{www}, Campus de la Universitat de les Illes Balears,
E-07122 Palma de Mallorca, Spain.}
\thanks[www]{\tt http://www.imedea.uib.es/PhysDept/}

\begin{abstract}
A model of interacting random walkers is presented and shown to give rise to
patterns consisting in periodic arrangements of fluctuating particle clusters.
The model represents biological individuals that die or reproduce at rates
depending on the number of neighbors within a given distance. We evaluate the
importance of the discrete and fluctuating character of this particle model on
the pattern forming process. To this end, a deterministic mean-field description,
including a linear stability and a weakly nonlinear analysis, is given and
compared with the particle model. The deterministic approach is shown to
reproduce some of the features of the discrete description, in particular, the
existence of a finite-wavelength instability. Stochasticity in the particle
dynamics, however, has strong effects in other important aspects such as the
parameter values at which pattern formation occurs, or the nature of the
homogeneous phase.

\end{abstract}

\date{November 30, 2003}
\end{frontmatter}


\section{Introduction}
\label{section:intro} Pattern formation in the presence of
fluctuations \cite{sancho,sanmiguel} is a subject attracting much
interest since the beginnings of instability studies. The most
common way to model the process is in terms of stochastic field
equations in which the deterministic part undergoes a pattern
forming instability and a noise term modifies the dynamics. In
situations where fluctuations are of {\sl internal} origin, as for
example arising from the discrete nature of the particles
integrating the system, a more fundamental description is in terms
of the microscopic particle dynamics (an analogy  between a
particular competing individual system with a reaction-diffusion
system is presented in \cite{bagnoli}). Models of interacting
particle systems leading to periodic pattern formation are however
not abundant in the literature. We introduced recently \cite{our}
one such a model, inspired in the interactions of biological
individuals that compete for the resources around them and
generalizing a  model introduced in the context of plankton
populations \cite{YoungNature} . It consists in a set of random
walkers, each one reproducing or dying with probabilities
depending on the number of other individuals in a neighborhood.
In our previous work \cite{our} we presented a detailed numerical study of the
model in two dimensions whose main result can be summarized in the existence of a
stationary periodic pattern of clusters of particles for a specific range of
parameters. A continuum Langevin equation is then derived under certain
approximations for the density of particles, with a rather complicated noise
term. A linear stability analysis is then performed on the deterministic
(noiseless) equation that in two dimensions predicts the transition to periodic
patters by a finite-wavelength instability. However, some numerical results, like
the value of the parameters at which patterns emerge, are not well reproduced in
the deterministic density equation. This reflects that the role of fluctuations
(the noise term in the Langevin equation) maybe rather important depending on the
properties of the system one wants to focus on.

In this context,  we present in this Paper additional results for the model, with
more emphasis on the one-dimensional case, and compare it more carefully with
descriptions in which fluctuations are absent: a mean-field approximation and its
amplitude equation. The onedimensional case is relevant by itself (see for
example a relevant experimental situations in \cite{lin} and references therein),
and also because it allows for a simpler analytical treatment. The amplitude
equation in this work is calculated for both the one- and two-dimensional
situations, but it is only the onedimensional expression which turns out to be
suitable for full comparisons with the numerics. It is also worth mentioning some
other extensions with respect to the results shown in \cite{our}: the analytical
results are stated for a general Kernel (see Sect. IV) of the mean-field density
equation, and two order parameters are identified in the model, $\sqrt{I(0)}$ and
$\sqrt{I(k_m)}$, which inform, respectively, about the empty-active transition
and the pattern formation transition (see Sect. III).


The Paper is organized as follows: In Sect. \ref{section:model} we present the
particle model, and show numerical results on its behavior in Sect.
\ref{section:numerical}. Section \ref{section:continuum} deals with the
mean-field approximation, including its linear and weakly nonlinear analysis. We
close with the Conclusions.

\section{A model of nonlocally interacting individuals}
\label{section:model}

We consider here both the one-dimensional situation ($d=1$), in which the system,
consisting of $N(t)$ particles, is contained in a line, and the two-dimensional
one ($d=2$), in which the $N(t)$ particles are in a square. In both cases we use
periodic boundary conditions. At the starting time $t=0$ an initial population of
$N(0)$ particles is randomly distributed. At time $t$, when the population is
$N(t)$, one particle is selected at random (say particle $j$) and it dies (thus
disappearing from the system) with probability $p_j$, reproduces (i.e. it
replicates itself) with probability $q_j$, or is left unchanged with probability
$1-p_j-q_j$. In the case of reproduction the newborn is placed at the same
location as the parent particle. The process is repeated $N(t)$ times. After
this, each particle moves independently in a random direction for a distance
drawn from a Gaussian distribution of standard deviation $\sigma$, then time is
incremented an amount $\tau$ ($t\rightarrow t+\tau$), and the algorithm repeats
with the resulting $N(t+\tau)$ individuals. The random motion leads to diffusion
with a diffusion coefficient $D=\sigma^2/2\tau$. The key point is the election of
the probabilities $p_j$ and $q_j$ or, equivalently, the death and reproduction
rates $\beta_j=p_j/\tau$ and $\lambda_j=q_j/\tau$, respectively. We choose a
constant death rate $\beta_j=\beta_0$, $\forall j$, but the reproduction rate of
particle $j$ depends of the number of particles $N_R^j$ within a distance $R$
from its position:
\BA
\lambda_j=\left\{
\begin{array}{cl}
\lambda_0-g N_R^j&\mbox{if} \ \  N_R^j  \le \lambda_0/g  \\
0&\mbox{if} \ \ N_R^j \ge \lambda_0/g
\label{ndr}
\end{array}\right.
\EA
This kind of interaction models a slowing down of birth rates (until total
suppression) in regions of high particle density. This is rather appropriate to
model biological populations that compete for food or other resources present in
a neighborhood of their positions, or that release toxic chemicals.

We choose $\lambda_0+\beta_0=1/\tau$, and introduce the maximum net growth rate
 $\mu\equiv \lambda_0-\beta_0$. For fixed $\tau$, the value of $\mu$ fixes both
$\lambda_0$ and $\beta_0$. Since $p_j$ and $q_j$ are probabilities, we see that
$\mu\tau \in [-1,1]$. One can estimate an equilibrium average density in this
model by imposing that at each site birth and death compensate on average:
$\lambda_0-g\langle N_R^j \rangle=\beta_0$. By assuming a uniform average
density, $\phi$, one can express the number of neighbors as $\langle N_R^j
\rangle=\phi h R^d$ ($h$ is $2$ in one dimension and $\pi$ in two dimensions, so
that $hR^d$ is the length or the area of the neighborhood of a particle; we have
not discounted the central particle from the number $N_R^j$, since a difference
of one particle is irrelevant in the regime this argument is intended to
describe). Thus
\BE
\phi \approx \frac{\mu}{ghR^d} \ .
\label{homog-discrt}
\EE

But this estimation completely neglects inhomogeneities and diffusion.
Fluctuations are known to produce inhomogeneities that alter estimations such as
(\ref{homog-discrt}). This is specially true in situations like this where, in
addition to (\ref{homog-discrt}), there is another homogeneous state where birth
and death processes also compensate: the empty state $\phi=0$ ($N(t)=0$), for
which both rates vanish. This is an absorbing state \cite{mamunoz} since further
evolution is impossible if the system reaches it. A phase transition to the
absorbing state is expected by reducing the net growth rate $\mu$ from a higher
value. The character of the transition and the location of the transition point
are not well reproduced if fluctuations are neglected
\cite{mamunoz,Hinrichsen2000}.

\section{Numerical observations}
\label{section:numerical}

\begin{center}
\begin{figure}
\epsfig{file=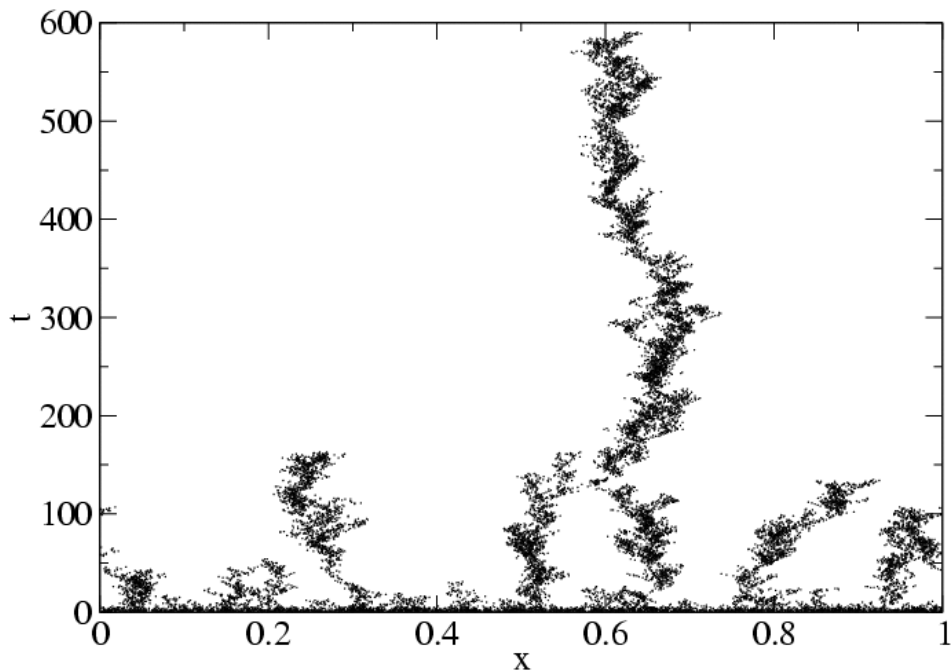,width=0.9\linewidth}
\caption{Time evolution (in vertical) of the particle positions (in horizontal)
for $\mu=0.4$, $D=10^{-5}$, $R=0.1$, and $g=1/50$, starting from an initial
population of 1000 particles. Units of time are such that $\tau=1$, and system
size is $L=1$. The final state is empty of particles}
\label{fig:absorbing}
\end{figure}
\end{center}

We first consider the one-dimensional case. The particle model described above is
run on a ring of length $L=1$. For $\mu$ smaller than a critical value $\mu_0$,
all initial particle distributions that we have checked end up attracted by the
absorbing empty state. An illustrative way of showing one-dimensional particle
dynamics consists in plotting the particle positions in the horizontal axis and
the time evolution of these in the vertical. Proceeding in this way, Fig.
\ref{fig:absorbing} shows an example of the dynamics of the emptying process,
strongly reminiscent of directed percolation \cite{Hinrichsen2000} below
threshold.

\begin{center}
\begin{figure}
\epsfig{file=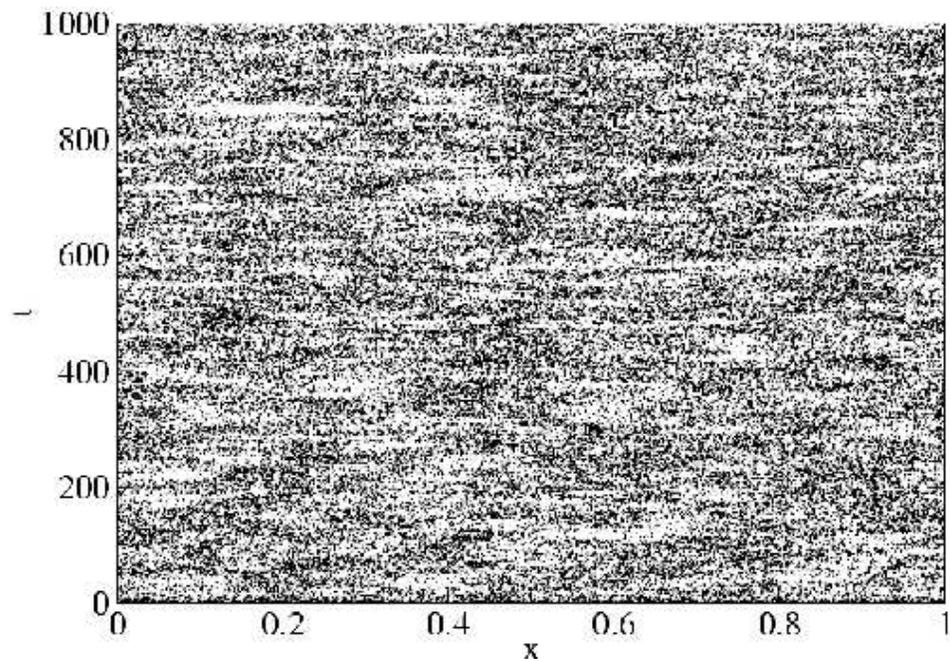,width=0.9\linewidth}
\caption{Idem as in Fig. (\ref{fig:absorbing}) except for $\mu=0.46$ and
$D=3\cdot 10^{-4}$. An active homogeneous phase has developed. }
\label{fig:uniform}
\end{figure}
\end{center}

By increasing the net growth, $\mu$, two situations appear: If the diffusion
coefficient is large enough, an active phase, roughly homogeneous as shown in
Fig. \ref{fig:uniform}, appears above a critical value $\mu_0$. The value of
$\mu_0$ is larger than the one required to equilibrate maximum birth rate and
deaths ($\mu=0$). Moreover the average density is smaller than the predicted by
the crude estimation (\ref{homog-discrt}).

\begin{center}
\begin{figure}
\epsfig{file=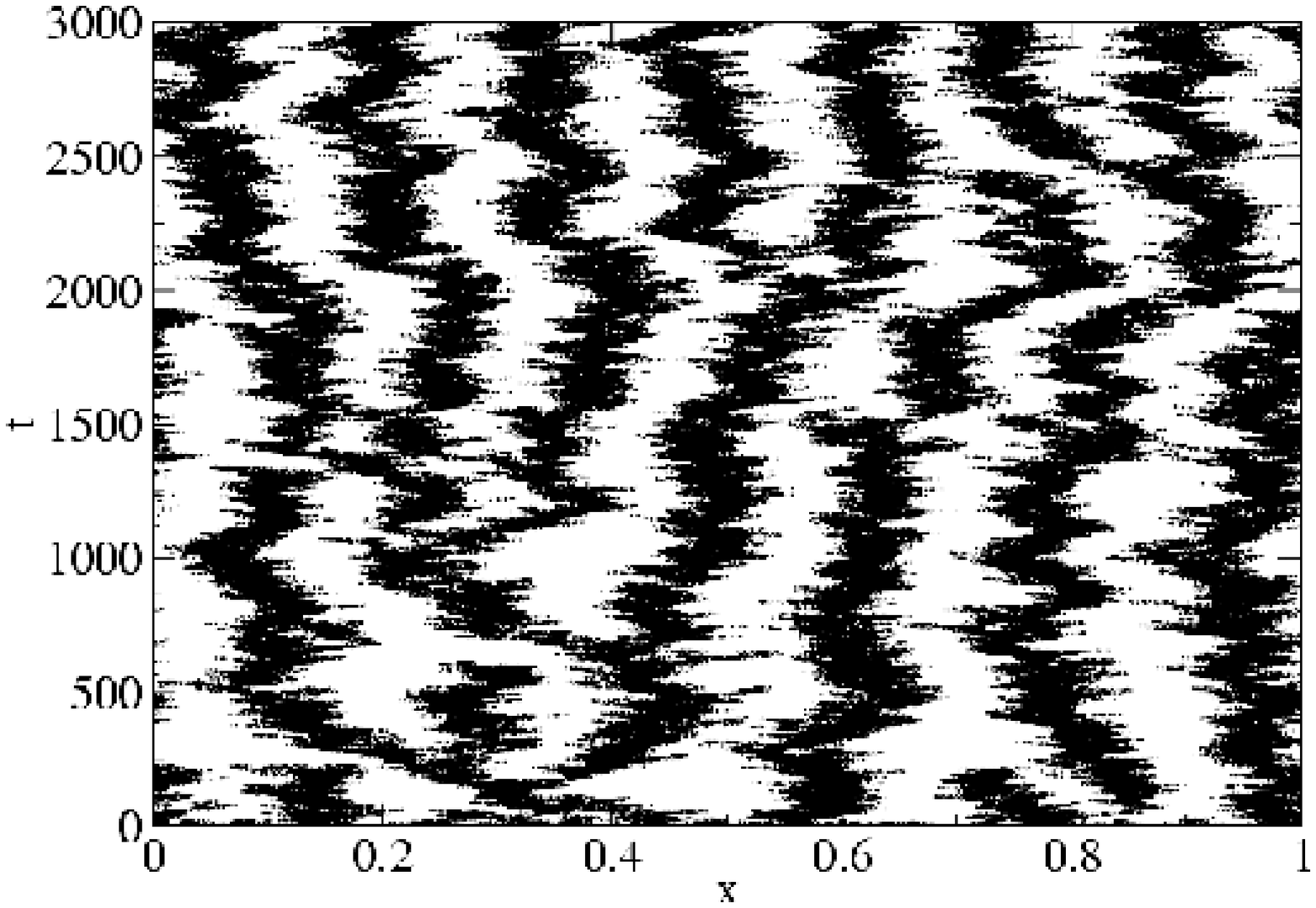,width=0.9\linewidth}
\caption{Idem as in Fig. (\ref{fig:absorbing}) except for $\mu=0.50$. Noisy pattern
formation has occurred.}
\label{fig:noisy}
\end{figure}
\end{center}

\begin{center}
\begin{figure}
\epsfig{file=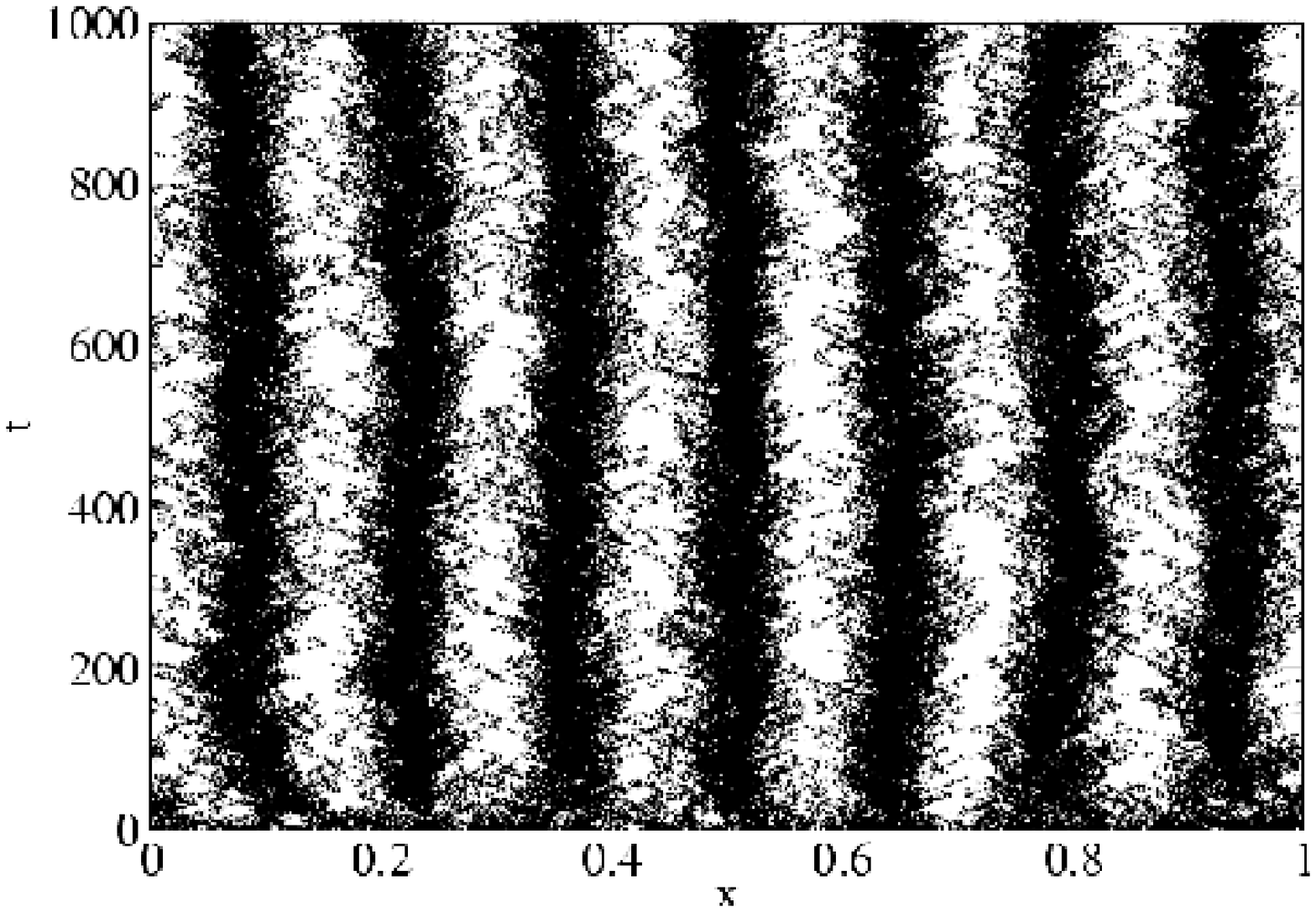,width=0.9\linewidth}
\caption{Idem as in Fig. (\ref{fig:absorbing}) except for $\mu=0.90$. A rather well
defined pattern is reached soon.}
\label{fig:pat}
\end{figure}
\end{center}

The other situation appears for smaller diffusion coefficients: now increasing
$\mu$ above $\mu_0>0$ particles organize in clusters which are roughly
equidistant, so that a periodic pattern results (Figs. \ref{fig:noisy} and
\ref{fig:pat}). The configurations are rather noisy, specially close to the
transition (Fig. \ref{fig:noisy}), but a periodicity in the pattern is clear.

\begin{center}
\begin{figure}
\epsfig{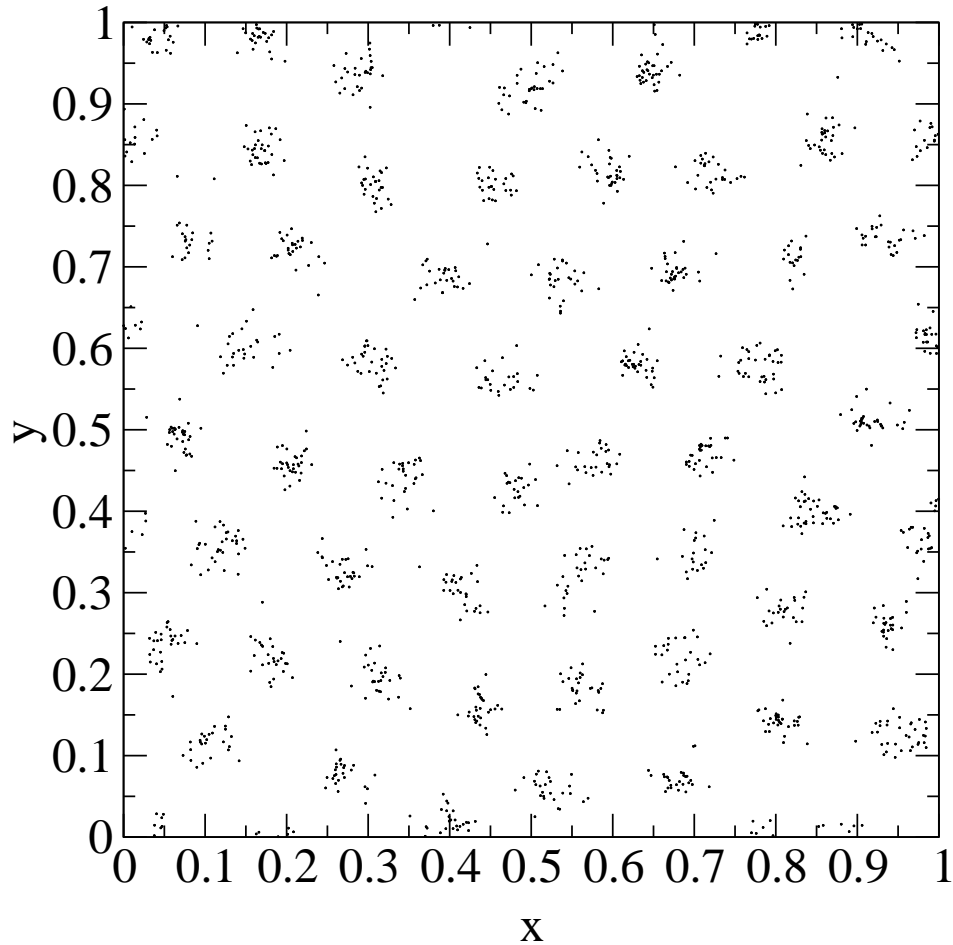}
\caption{A snapshot of the fluctuating two-dimensional pattern of particle clusters,
obtained in a square of side $L=1$ for $\mu=0.7$, $D=10^{-5}$, $R=0.1$, $g=1/50$,
and $\tau=1$.}
\label{fig:disc2d}
\end{figure}
\end{center}

In the two-dimensional case the general scenario is similar, and was already
described in \cite{our}. Patterns consist in fluctuating particle clusters
arranged in a hexagonal lattice. We show a snapshot in Figure \ref{fig:disc2d}.

In order to be more quantitative, we introduce a structure function
\BE
I(k) = \left< \left|   \sum_{j=1}^{N(t)} e^{i \bk \cdot \bx_j} \right|^2 \right>
\ .
\label{structure}
\EE

\begin{center}
\begin{figure}
\epsfig{file=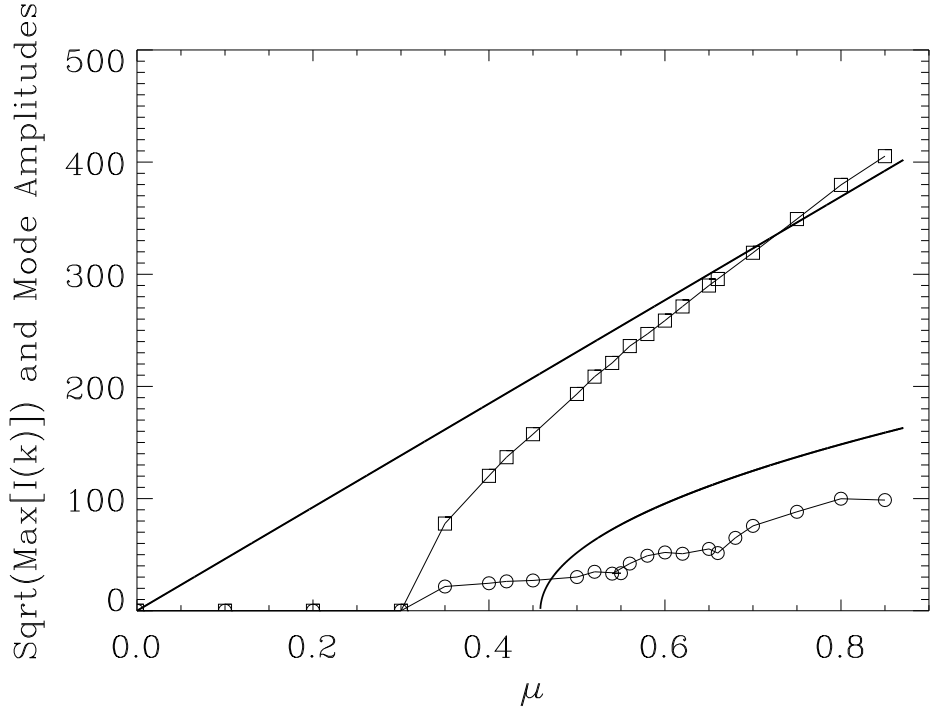,width=\linewidth}
\caption{Values of $\sqrt{I(0)}$ (squares) and $\sqrt{I(k_M)}$ (circles) as a function
of $\mu$ for the one-dimensional particle model at $D=9.2\cdot 10^{-5}$,
$R=0.13$, $g=1/120$, $\tau=1$ and $L=1$, for which $k_M=10\pi$. The upper thick
line is $\phi_s$, from (\ref{phis1d}), and the lower one is $\phi_p$, from
(\ref{phip}).}
\label{fig:maximos}
\end{figure}
\end{center}

The sum is over particles, at positions $\bx_j$, and the average is a temporal
average on a long-time state, performed to improve statistics. Maxima in $I(k)$
identify the wavenumbers, $k=|\bk|$, associated to periodicities in the system.
In any nonempty state there is also a peak in $k=0$ giving the square of the mean
particle density. The values $I(0)$ and $I(k_M)$, where $k_M$ is the nonvanishing
wavenumber at which there is a maximum, act as two order parameters: the first
one for the transition from the empty to the active phase, and the second for the
pattern formation. For Figs. \ref{fig:noisy} and \ref{fig:pat}, $k_M=14\pi$,
corresponding to the observation that the pattern consists of $7$ cells (a
pattern of $n$ cells is characterized by a wavenumber $k_M=2\pi n/L$). In Fig.
\ref{fig:maximos} we plot the value of  $\sqrt{I(0)}$ and $\sqrt{I(k_M)}$ as a
function of $\mu$ in a different case for which $k_M=10\pi$ (i.e. 5 cells). Both
order parameters become nonvanishing at the same absorbing transition, although
with rather different amplitudes and behavior.

\section{Mean-field description}
\label{section:continuum}

We try now to understand the pattern forming processes described in the previous
Section. It is obvious that the patterns are rather noisy, suggesting that
fluctuations play a r{\^o}le in its development. It is known from related models
\cite{Zhang90,YoungNature,Adler97} that, in addition to its impact on the
absorbing transition, reproductive fluctuations have another important effect
arising from the asymmetry between birth and death: Birth is a multiplicative
process that increments the density in regions where it is already high, whereas
death can occur anywhere. Thus, even if the average rates are equal, under
reproductive fluctuations there is a tendency of the particles to organize into
clusters of high density, leading to very inhomogeneous configurations when
diffusion is not strong and is the only homogenizing force
\cite{Zhang90,YoungNature,Adler97}. This inhomogeneity is amplified when birth
rate exceeds death, but it is not observed in the common situations in which the
growth is saturated by local interactions \cite{Hinrichsen2000}. Thus, it is not
obvious whether the pattern development in our model arises from the stochastic
microscopic fluctuations, or if rather it is caused by some deterministic
instability arising from the nonlocal interactions.

To address this question we can write down a mean-field-like description of the
model, which completely neglects fluctuations, and check if the instability is
present there. The mean-field equation is written in terms of an {\sl expected
density} $\phi(\bx,t)$ as follows
\BE
\partial_t \phi(\bx,t)=\mu \phi(\bx,t)+D\nabla^2\phi(\bx,t)-
g\phi(\bx,t)\int_S G(\bx-\by) \phi(\by,t) d\by \ .
\label{continuum}
\EE
The first and second terms are the standard mean-field description of the net
growth and the diffusion processes. The last one is a nonlocal contribution
associated to the saturation produced by the neighborhood interactions in our
model if the kernel $G(\bx)$ is given by
\BA
G(\bx)= \left\{
\begin{array}{cl}
1&\mbox{if}\ \ | \bx | \le R   \\
0&\mbox{if}\ \ | \bx | \ge R.
\end{array}\right.
\label{kernel}
\EA
$S$ is the domain were the system is defined. Equation (\ref{continuum}) was more
systematically deduced in \cite{our} from our microscopic model, as the
deterministic part of a Langevin equation containing rather complex noise terms.
Equation (\ref{continuum}) and related ones have also been proposed independently
to model directly the macroscopic behavior of a variety of biological systems
\cite{FuentesPRL,FuentesPre,JTB,Shnerb}. In these cases, other kernels were
considered in addition to (\ref{kernel}), and for some of them pattern formation
was observed. In the following we establish properties of (\ref{continuum}) for
general interaction kernels such that $G=G(|\bx|)$, although the  final results
will be only commented for the step kernel given by (\ref{kernel}). We first
establish general relationships, then perform a linear stability analysis of a
homogeneous solution, and then consider the weakly nonlinear behavior. We
restrict to domains $S$ either unbounded or with periodic boundary conditions.

\subsection{General relationships and homogeneous solutions}

It is useful to write Eq. (\ref{continuum}) in non-dimensional form to have a
clearer idea of the independent parameters involved. We rescale the units of
space, time, and $\phi$ as follows:
\BE
\bu =\frac{\bx}{R}\ , \ \ \  s = \frac{t}{R^2/D}\ ,  \ \ \
\Psi(\bu,s)=\frac{\phi(\bx,t)}{D/gR^{2+d}}  \ .
\label{scaling}
\EE
In this way:
\BE
\partial_t \Psi(\bu,s)=\tilde \mu \Psi(\bu,t)+\nabla_u^2\Psi(\bu,t)-
\Psi(\bu,t)\int_E H(\bu-\bv) \Psi(\bu,t) d\by \ .
\label{adim}
\EE
$E$ is the rescaled domain, and $H(\bu)$ is the rescaled kernel:
$H(\bu)=G(\bx=R\bu)$. In this mean-field description, The only relevant parameter
turns out to be
\BE
\tilde \mu \equiv \frac{\mu R^2}{D} \ .
\label{nu}
\EE
For future reference we introduce
\BE \hat H (q)=\int_E  H(\bu) e^{i\bq\cdot\bu} d\bu    \ ,
\EE
and the notation $h \equiv \hat H(0)=\int_E  H(\bu) d\bu$. Because of the
symmetry assumed for $G(\bx)$, $\hat H (q)$ depends only on the modulus of the
wavenumber $q=|\bq|$. For our step kernel we have, in the one-dimensional case:
\BE
\hat H(q)=2 \frac{\sin(q)}{q} \ , \ \ \ h=2
\label{sinc}
\EE and for the
two-dimensional case
\BE
\hat H(q)=2\pi  \frac{J_1(q)}{q} \ , \ \ \ h=\pi  \ .
\label{bessel}
\EE
$J_1(q)$ is the first order Bessel function.

The trivial solution $\Psi$=$0$ of (\ref{adim}) becomes unstable for
$\tilde\mu>0$. Then a uniform solution appears:
\BE
\Psi(\bu,t)=\frac{\tilde \mu}{h} \equiv \nu \ ,
\label{homosolution}
\EE
corresponding to
\BE
\phi(\bx,t)=\frac{\mu}{g h R^d} \equiv \phi_s\ .
\EE
This reproduces the crude approximation (\ref{homog-discrt}), performed for the
particle model. It is convenient to introduce the deviation $\psi$ with respect
to the homogeneous solution:
\BE
\Psi(\bu,t)=\frac{\tilde \mu}{h}+\psi(\bu,t) \ ,
\label{shifthomosolution}
\EE
so that (\ref{adim}) reads
\BA
\partial_t\psi(\bu,t) &=& \nabla_u^2 \psi(\bu,t)      \nonumber \\
&-& \nu \int_E H(\bu-\bv) \psi(\bv,t) d\bv - \psi(\bu,t) \int_E H(\bu-\bv)
\psi(\bv,t) d\bv \ .
\label{shiftedequation}
\EA

\subsection{Linear stability analysis}

\begin{center}
\begin{figure}
\epsfig{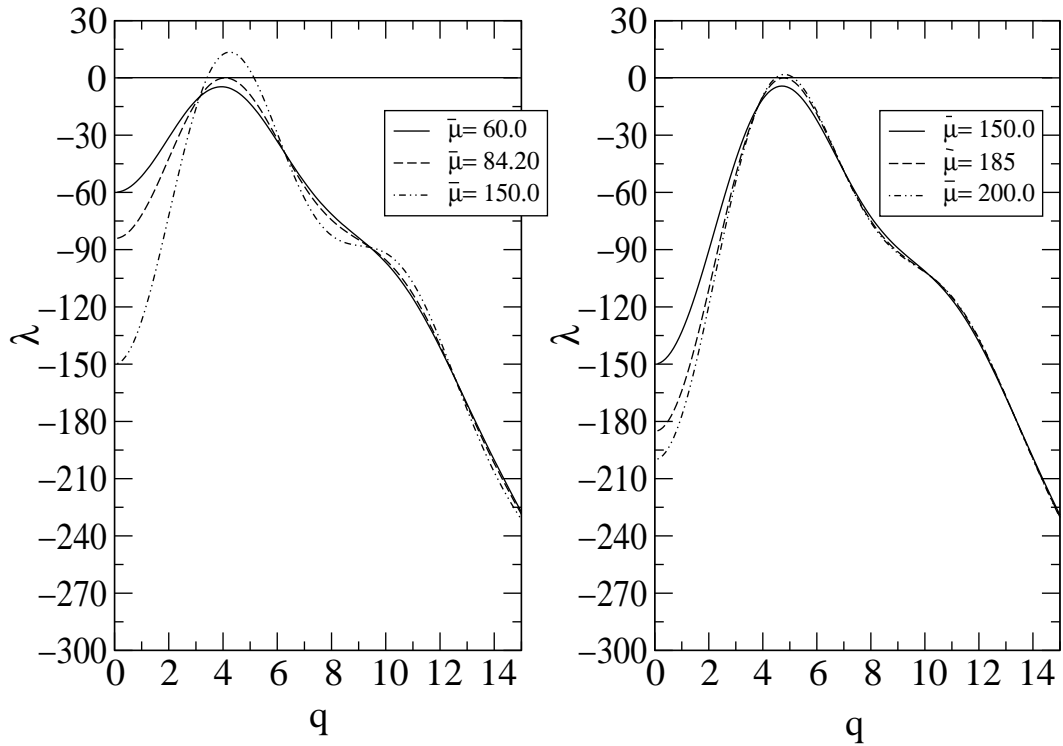}
\caption{The linear
growth rate (\ref{lambdaq}) of perturbations of wavenumber $q$ to the homogeneous
solution (\ref{homosolution}). Left: $d=1$. Right: $d=2$. In both cases positive
growth, i.e. instability, arises by increasing $\tilde \mu$.}
\label{fig:dispersion}
\end{figure}
\end{center}

\begin{center}
\begin{figure}
\epsfig{file=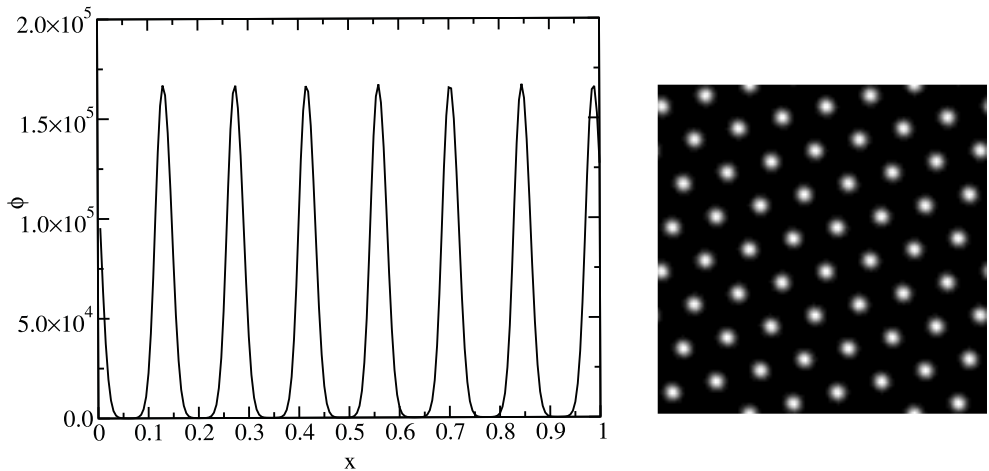,width=\linewidth}
\caption{Steady periodic patterns obtained by long-time integration of the mean-field
model (\ref{continuum}) in $d=1$ (left) and $d=2$ (right). In both cases
$\mu=0.7$, $D=10^{-5}$, $R=1$, and $g=1/50$. In the second case higher values of
$\phi$ are represented by lighter gray levels. }
\label{fig:DetPat}
\end{figure}
\end{center}

The linear stability analysis of the homogeneous solution can now
readily performed by introducing perturbations of the form $\psi
\approx \exp(i\bq\cdot\bu+\lambda(k)t)$ and neglecting nonlinear
terms. The perturbation growth rate is
 \BE \lambda(q)=-q^2-\nu
\hat H(q) \ . \label{lambdaq}
\EE
 It is seen that a necessary
condition for instability is that $\hat H(q)$ takes negative
values (note that Eq. (\ref{lambdaq}) generalizes to an arbitrary
kernel the expression shown in \cite{our} for the step kernel) .
It is also sufficient, since instability will always be reached in
this case by increasing sufficiently $\nu$ (or $\tilde \mu$).
Thus, if $\hat H(q)<0$ for some range of $q$, there is a critical
value of $\nu$, $\nu_c$ (corresponding to a critical
$\tilde\mu_c$) such that an instability to pattern formation
occurs \cite{FuentesPre,our}. Since both (\ref{sinc}) and
(\ref{bessel}) take negative values, we see that the mean-field
model with the step kernel undergoes a pattern forming instability
without the need of any stochastic noise, as already reported
\cite{FuentesPRL,our,FuentesPre}. Figure \ref{fig:dispersion}
(left) plots expression (\ref{lambdaq}) in the one-dimensional
case, showing the change of sign of $\lambda(q)$ around a critical
$q_c$ by increasing $\tilde \mu=\nu h$. Figure \ref{fig:DetPat}
(left) shows a long-time numerical solution of
(\ref{continuum})-(\ref{kernel}) in one dimension, displaying a
well-developed steady pattern. The right part of Fig.
\ref{fig:DetPat} shows the analogous structure in two dimensions:
the instability leads to the formation of a pattern of hexagonal
symmetry, as in the particle model. The corresponding
two-dimensional dispersion relation (\ref{lambdaq}) is shown in
the right part of Fig. \ref{fig:dispersion}.

Eq. (\ref{lambdaq}) explains also the absence of instability for other kinds of
kernels, such as the Gaussian one, for which $\hat H(q)$ is positive definite
\cite{JTB,FuentesPRL,FuentesPre} (we stress however that patterns may appear
under front propagation, or in response to spatial inhomogeneities \cite{JTB}).
To make a more quantitative analysis, the value of $\nu_c$ and the fastest
growing wavenumber $q_c$ are obtained from the simultaneous solution of
\BA
q_c^2+\nu_c\hat H(q_c) &=& 0   \nonumber \\
2q_c+\nu_c\hat H'(q_c) &=& 0 \ .
\EA
Numerical solution of these equations leads, for the one-dimensional case and
step kernel:
\BE
q_c \approx 4.078  \ , \ \ \ \nu_c \approx 42.10 \ ,
\label{qcnuc1d}
\EE
so that $\tilde \mu_c\approx 84.20$. In two dimensions:
\BE
q_c \approx 4.779 \ , \ \ \ \nu_c \approx 58.948 \ ,
\label{qcnuc2d}
\EE
and $\tilde \mu_c\approx 185.192$. From (\ref{scaling}), the unscaled values
$\bk$ of the wavenumbers are given by $\bk=\bq/R$.

These predictions can be compared with the numerical observations of the particle
model. First, the maximum of the structure function for the patterns in the Figs.
\ref{fig:noisy} and \ref{fig:pat} was at $k=14 \pi \approx 43.98$. This is
precisely the first wavenumber compatible with the periodic boundary conditions
immediately above the fastest growing mode in the mean-field equation for
$R=0.1$: $k_c=q_c/R \approx 40.78$. For the configurations of periodic character
among the ones used to construct Fig. \ref{fig:maximos}, the maximum in the
structure function was at $k=10 \pi \approx 31.4159 $, to be compared with
$k_c=q_c/R\approx 31.369$ for $R=0.13$. The agreement is thus excellent in both
cases. We can conclude that the essential mechanism of the pattern forming
instability is well captured by the deterministic description, being the period
of the pattern even quantitatively reproduced. For the two-dimensional case, the
same agreement was already observed \cite{our}.

The situation is different for the location of the point of pattern onset. As for
the location of the absorbing transition, the mean-field prediction and the
particle model disagree. Typically, the transition from the empty state to the
pattern is observed directly, without the intermediate of the homogeneous
distribution. Only if diffusion is greatly increased the homogeneous state is
observed (Fig. \ref{fig:uniform}) but then the transition to patterns predicted
by (\ref{qcnuc1d}) occurs at $\mu$ exceeding the maximum of the physical range
$\mu\tau \in [-1,1]$ (and it is indeed not observed in the particle model, which
is a kind of agreement). The reason for this disagreement, despite containing the
correct instability mechanism, is easy to understand: In cases such as the ones
in Sect. \ref{section:numerical} with small $D$ ($D=10^{-5}$) it happens that the
predicted value of $\mu_c$ ($=0.084$) is much {\sl smaller} than the $\mu_0$ of
the absorbing transition (which is $\approx 0.45$). Thus it is not strange than,
as soon as the active phase becomes preferred above $\mu_0$, it is already of the
periodic type. The failures of the mean-field approach to fully describe the
pattern behavior are then due to the strong impact of microscopic fluctuations on
the absorbing transition. A second case is the one corresponding to Fig.
\ref{fig:maximos}. Here $\mu_c=0.458 > \mu_0 \approx 0.3$. But what is observed
is that, even when the system is below threshold, microscopic fluctuations excite
the less-damped modes \cite{sancho,sanmiguel}, so that a small amplitude noisy
pattern is always present above the absorbing transition, as seen in Fig.
\ref{fig:maximos}. Fluctuation strength is relatively large here because of the
low amplitude of the active state close to $\mu_c$, caused by the proximity to
$\mu_0$. Pattern formation is not a sharp bifurcation in this case, but a
continuous crossover. Finally, in the situation in which the absorbing transition
and the pattern forming instability would be well separated, this last one falls
outside the physical range $\mu\tau \in [-1,1]$.

\subsection{Weakly nonlinear amplitude equations}

The advantage of having a continuum model as (\ref{continuum}) is the possibility
to use the tools of pattern formation theory \cite{patterns,Walgraef} to obtain
analytic results. A first example has been the linear stability analysis
performed before. We now estimate the pattern amplitude close to threshold by
calculating the pertinent amplitude equations.

To this end we start from Eq. (\ref{shiftedequation}), and perform the standard
expansions in a small parameter $\epsilon$ related to the distance to threshold:
\BE
\psi(\bx,t)=\epsilon\psi_1(\bu,T_1,T_2,
...)+\epsilon^2\psi_2(\bu,t,T_1,T_2,...)+\epsilon^3\psi_3(x\bu,t,T_1,T_2,...)+
...
\EE
\BE
\nu=\nu_c+\epsilon\nu_1+\epsilon^2\nu_2+ ...
\EE
We have introduced two slow time scales: $T_1=\epsilon t$, $T_2=\epsilon^2 t$.
For simplicity we do not introduce additional space variables, needed when
considering large-scale pattern modulations.

To first order in $\epsilon$, one finds the equation
\BE
L_c\psi_1(\bu,T_1,T_2)=0 \ , \ \ {\rm with}\ \ L_c=\partial_t-\nabla_u^2+\nu_c
\HH \ \ .
\label{critical}
\EE
$\HH$ is the integral operator acting as $\HH f(\bu) \equiv \int_E
H(\bu-\bv)f(\bv)d\bv$. Eq. (\ref{critical}) implies that $\psi_1$ is a
combination of periodic patterns of the critical periodicity $2\pi/q_c$:
\BE
\psi_1(\bu,T_1,T_2)=\sum_j A_j(T_1,T_2) e^{i \bq_j \cdot \bu} \ ,
\label{psi1}
\EE
with $|\bq_j|=q_c$.

At orders $\epsilon^2$ and $\epsilon^3$ we find
\BA
L_c\psi_2 &=& -\partial_{T_1}\psi_1-\nu_1\HH \psi_1-\psi_1\HH\psi_1
\label{eps2}\\
L_c\psi_2 &=&-\partial_{T_1}\psi_2-\partial_{T_2}\psi_1 - \nu_1 \HH \psi_2 -
\nu_2 \HH \psi_1 - \psi_2\HH\psi_1-\psi_1\HH\psi_2 \ \ .
\label{eps3}
\EA

We particularize to the one-dimensional case, in which the only possible planform
of the form (\ref{psi1}) is
\BE
\psi_1(u,T_1,T_2)= A(T_1,T_2) e^{i q_c u}+A^*(T_1,T_2) e^{-i q_c u} \ .
\label{rolls}
\EE
$A^*$ is the complex conjugate of $A$. Substituting in (\ref{eps2}) one finds, as
usual in one-dimensional patterns, that boundedness of the solutions implies
$\nu_1=0$, so that $\epsilon \propto (\nu - \nu_c)^{1/2}$ and
$A(T_1,T_2)=A(T_2)$. By solving (\ref{eps2}) and substituting in (\ref{eps3}),
one finds the condition to avoid resonant terms:
\BE
\partial_{T_2} A = -\nu_2 \hat H(q_c) A - \kappa |A|^2 A \ ,
\label{AamplitudeEq}
\EE
which is the desired amplitude equation. Here,
\BA
\kappa &=&\hat H(q_c) (Q+S)+\hat H(2q_c) Q+\hat H(0)S  \nonumber\\
Q &=& \frac{-\hat H(q_c)}{4q_c^2+\nu_c\hat H(2q_c)} \ \ \ ,\ \ S = \frac{-2\hat
H(q_c)}{\nu_c\hat H(0)}
\EA
All these quantities are pure non-dimensional numbers. Their values, in addition
to the ones given in (\ref{qcnuc1d}), are $\hat H(0)=h=2$, $\hat H(q_c) \approx
-0.3950$, $\hat H(2q_c) \approx 0.2341$, $Q=5.172\cdot 10^{-3}$, $S=9.383\cdot
10^{-3}$, and $\kappa=0.01423$. The content of (\ref{AamplitudeEq}) is more
clearly seen by reintroducing the original time variable $t=\epsilon^{-2}T_2$ and
the amplitude $B(t)=\epsilon A(T_2)$ so that
\BE
\dot B(t) = (\nu-\nu_c)\left( -\hat H(q_c) \right) B - \kappa |B|^2 B
\label{BamplitudeEq}
\EE
The equation predicts that instability develops for $\nu>\nu_c$ and leads to a
pattern of arbitrary phase, and amplitude
\BE
|B|=\sqrt{  (\nu-\nu_c)\frac{-\hat H(q_c)}{\kappa}  }
\label{steady}
\EE
that has bifurcated supercritically from the homogeneous solution
(\ref{homosolution}). Returning back to dimensional variables via
(\ref{scaling}), the periodic steady solution for the original density can be
reconstructed as
\BE
\phi(x,t) = \phi_s   + \phi_p  \left(   e^{i (k_c x+\varphi)} + e^{-i (k_c
x+\varphi)} \right)+ {\mathcal O}(\mu-\mu_c) \ .
\label{final}
\EE
with
\BE
\phi_s=\frac{\mu}{2gR}
\label{phis1d}
\EE
and
\BE \phi_p = \sqrt{\mu-\mu_c}\ \frac{D^{\frac{1}{2}}}{gR^2}\ \sqrt{\frac{\hat
H(q_c)}{2\kappa}} \ .
\label{phip}
\EE
$\varphi$ is an arbitrary phase. We can compare this analytic prediction  with
the particle patterns. The structure function (\ref{structure}) can be
interpreted as the power spectrum of a density of the form $\sum_{i=1}^{N(t)}
\delta(\bx-\bx_i)$. Comparing with the power spectrum of the mean-field density
(\ref{final}), we see that $\sqrt{I(0)}$ should be related to $\phi_s$, and
$\sqrt{I(k_M)}$ to $\phi_P$. We remind that $k_M$ was well reproduced by the
mean-field value of $k_c$. Comparison is performed in Fig. \ref{fig:maximos}. We
see that the agreement in the amplitudes is poor (even if it was expected to be
valid only close to $\mu_c$) although some of the trends are reproduced. The
disagreement is linked to the failure to capture the correct transition points.

The amplitude equation can also be calculated in $d=2$, by using an hexagonal
planform \cite{patterns,Walgraef} instead of (\ref{rolls}):
\BE \psi_1(\bu,T_1,T_2)= \sum_{j=1,2,3} A_j(T_1,T_2) e^{i \bq_j \cdot u}+ {\rm c.c.}
\label{hexagons}
\EE
$\bq_j$ ($j=1,2,3$), are three wavenumbers of modulus $q_c$ at $120$ degrees from
each other, and c.c. means complex conjugate.  The problem with such calculation
is that generically the bifurcation to hexagons is subcritical, so that there is
no guarantee that the amplitude equation will capture the saturation that occurs
far from the instability point. By using (\ref{eps2})-(\ref{eps3}) one finds,
\BE
\dot B_1(t)=(\nu-\nu_c)\left(-\hat H(q_c)\right) B_1 + 2\left(-\hat
H(q_c)\right)B_2^*B_3+{\mathcal O}(|B_j|^2B_1) \ ,
\label{HexAmplitudeEq}
\EE
and two other equations obtained by permuting cyclicly the indices $1,2,3$.
As in (\ref{BamplitudeEq}), $B_i$ is the coefficient of the Fourier mode of
wavenumber $\bq_i$ in the expansion of $\psi(\bx,t)$. Eq. (\ref{bessel}) should
be used for $\hat H(q_c)$. The sign of the third order terms not written
explicitly in (\ref{HexAmplitudeEq}) is such that they do not saturate the
pattern. Thus, expansion should be followed at higher orders and
(\ref{HexAmplitudeEq}) is of not much utility. Nevertheless, the presence of the
second order term confirms that the bifurcation in the mean-field model is
subcritical, and the positive sign of its coefficient (remember that $\hat
H(q_c)<0$) implies that the bifurcating solution leads to {\sl H0} or {\sl
positive} hexagons \cite{Walgraef} (i.e. points of high density forming an
hexagonal lattice), as observed in the particle model.

\section{Conclusion}
\label{section:conclusion}

In this work we have presented new results on a model recently introduced by the
authors \cite{our} which considers particles reproducing and dying at rates
depending on the number of individuals that every particle has in its $R$-ranged
neighborhood. The study has been performed at the level of the density equation
derived for this model, and mainly for the one-dimensional case, since analytical
results are simpler here.

As the general conclusion of the comparison between the particle and the
mean-field model we can say that, although the instability appears to be
deterministic in origin and the mean-field approach captures correctly the
wavenumber, fluctuations are important and affect transition points and
amplitudes. This limits the usefulness of standard pattern forming theory tools,
usually developed for continuum models as in the case of weakly nonlinear
analysis, and ask for further studies in which noise terms are explicitly
considered \cite{our}. Fluctuations may have also another interesting effect:
even in cases in which there is no deterministic instability, as for example when
a Gaussian kernel is used in (\ref{continuum}) \cite{JTB,FuentesPRL,FuentesPre},
it is very likely that a corresponding particle model would display a small
amplitude noisy pattern, arising by the less-damped models excited by
fluctuations.

\section*{Acknowledgments}
We acknowledge support from MCyT (Spain) projects BFM2000-1108
(CONOCE) and REN2001-0802-C02-01/MAR (IMAGEN). C.L. is a "Ram\'on
y Cajal" fellow of the Spanish MEC.

\end{document}